\documentclass[12pt]{article}

\usepackage{amsmath}
\allowdisplaybreaks
\usepackage{amssymb}

\usepackage{cancel}

\begin{document}

\baselineskip=16.7pt plus 0.2pt minus 0.1pt

\makeatletter
\@addtoreset{equation}{section}
\renewcommand{\theequation}{\thesection.\arabic{equation}}
\newcommand{\bm}[1]{\boldsymbol{#1}}
\newcommand{\calA}{\mathcal{A}}
\newcommand{\calB}{\mathcal{B}}
\newcommand{\calC}{\mathcal{C}}
\newcommand{\calE}{\mathcal{E}}
\newcommand{\calP}{\mathcal{P}}
\newcommand{\calM}{\mathcal{M}}
\newcommand{\calN}{\mathcal{N}}
\newcommand{\calV}{\mathcal{V}}
\newcommand{\calK}{\mathcal{K}}
\newcommand{\calF}{\mathcal{F}}
\newcommand{\calG}{\mathcal{G}}
\newcommand{\calH}{\mathcal{H}}
\newcommand{\calT}{\mathcal{T}}
\newcommand{\calU}{\mathcal{U}}
\newcommand{\calY}{\mathcal{Y}}
\newcommand{\calW}{\mathcal{W}}
\newcommand{\calL}{\mathcal{L}}
\newcommand{\calD}{\mathcal{D}}
\newcommand{\calO}{\mathcal{O}}
\newcommand{\calI}{\mathcal{I}}
\newcommand{\calQ}{\mathcal{Q}}
\newcommand{\calS}{\mathcal{S}}
\newcommand{\QB}{Q_\textrm{B}}
\newcommand{\nn}{\nonumber}
\newcommand{\drv}[2]{\frac{d #1}{d#2}}
\newcommand{\veps}{\varepsilon}
\newcommand{\eps}{\epsilon}
\newcommand{\ds}{\displaystyle}
\newcommand{\Ke}{K_{\veps}}
\newcommand{\invKe}{\frac{1}{\Ke}}
\newcommand{\Ue}{U_{\veps }}
\newcommand{\Ge}{G_{\veps }}
\newcommand{\Psie}{\Psi_{\veps}}
\newcommand{\CR}[2]{\left[#1,#2\right]}
\newcommand{\ACR}[2]{\left\{#1,#2\right\}}
\newcommand{\Pmatrix}[1]{\begin{pmatrix} #1 \end{pmatrix}}
\newcommand{\sPmatrix}[1]{
  \left(\begin{smallmatrix} #1 \end{smallmatrix}\right)}
\newcommand{\tr}{\mathop{\rm tr}}
\newcommand{\Tr}{\mathop{\rm Tr}}
\newcommand{\p}{\partial}
\newcommand{\wh}[1]{\widehat{#1}}
\newcommand{\wt}[1]{\widetilde{#1}}
\newcommand{\ol}[1]{\overline{#1}}
\newcommand{\abs}[1]{\left| #1\right|}
\newcommand{\VEV}[1]{\left\langle #1\right\rangle}
\newcommand{\Drv}[2]{\frac{\p #1}{\p #2}}
\newcommand{\ket}[1]{| #1 \rangle}
\newcommand{\bra}[1]{\langle #1 |}
\newcommand{\braket}[2]{\langle #1 | #2 \rangle}
\newcommand{\KBc}{K\!Bc}
\newcommand{\Ngh}{N_\textrm{gh}}
\newcommand{\Gm}{\Gamma}
\newcommand{\pF}{\calF}
\newcommand{\id}{\mathbb{I}}
\renewcommand{\Re}{\mathop{\rm Re}}
\renewcommand{\Im}{\mathop{\rm Im}}
\newcommand{\Res}{\mathop{\rm Res}}
\newcommand{\Ds}{\Delta_s}
\newcommand{\da}{\frac{2\pi i}{s}}
\newcommand{\Da}{\frac{2\pi i}{s}}
\newcommand{\opz}[1]{\left(1+z #1\right)}
\newcommand{\wpopz}[1]{\left(w+1+z #1\right)}
\newcommand{\whwpz}[1]{\left(\wh{z}+ #1\right)}
\newcommand{\CRBc}{{\mathcal X}}
\newcommand{\setalpha}{\{\alpha_k\}}
\newcommand{\Ld}{\Lambda}
\newcommand{\Vgrav}{\calV_{\textrm{grav.}}}
\newcommand{\calNgrav}{\calN_{\textrm{grav}}}
\newcommand{\Vmid}{\calV_{\textrm{mid}}}
\newcommand{\dg}{\ddagger}

\makeatother
\begin{titlepage}

\title{
\hfill\parbox{3cm}{\normalsize KUNS-2769}\\[1cm]
{\Large\bf
  Bernoulli Numbers and Multi-brane Solutions\\
  in Cubic String Field Theory
}}

\author{
Hiroyuki {\sc Hata}\footnote{
{\tt hata@gauge.scphys.kyoto-u.ac.jp}}
\\[7mm]
{\it
Department of Physics, Kyoto University, Kyoto 606-8502, Japan
}
}

\date{{\normalsize August 2019}}
\maketitle

\begin{abstract}
\normalsize

In a previous paper [arXiv:1901.01681], we presented an analytic
construction of multi-brane solutions in cubic open string field
theory (CSFT) for any integer brane number.
Our $(N+1)$-brane solution is given in the pure-gauge form
$\Psi=U\QB U^{-1}$ in terms of a unitary string field $U$ which is
specified by $[N/2]$ independent real parameters $\alpha_k$.
We saw that, for various sample values of $N$
$(=2, 3, 4, 5,\cdots)$, $\alpha_k$ can be consistently determined
by two requirements: The energy density from the action should
reproduce that of $(N+1)$-branes, and the EOM of the solution against
the solution itself should hold.

In this paper, we complete our construction by determining $\alpha_k$
satisfying the two requirements for a generic $N$.
We find that each $\alpha_k$ is given in a closed form by using the
Bernoulli numbers.
We also present some supplementary results on our solution;
the energy density of the solutions determined from its gravitational
coupling, and the unitary string field $U$ as an exponential
function.
\end{abstract}

\thispagestyle{empty}
\end{titlepage}

\section{Introduction}

In \cite{ACMBSol}, we presented an analytic construction of
multi-brane solutions in cubic open string field theory (CSFT)
\cite{CSFT}. The action and the equation of motion (EOM) of CSFT are
given respectively by
\begin{equation}
S=\int\!\left(\frac12\Psi*\QB\Psi+\frac13\Psi^3\right) ,
\end{equation}
and
\begin{equation}
\QB\Psi+\Psi*\Psi=0 .
\label{eq:EOM}
\end{equation}
Our construction is based on the $\KBc$ algebra \cite{Okawa}, and we
considered translationally invariant configuration of the pure-gauge
type as a candidate solution:\footnote{
  In our construction, we use only $(K,B,c)$. In contrast, there is a
  construction of solutions which uses the boundary condition changing
  operators in addition to the elements of the $\KBc$ algebra
  \cite{ErlerMacc}.
}
\begin{equation}
\Psi=U\QB U^{-1} ,
\label{eq:Psi=UQBU^-1}
\end{equation}
where $U=U(K,B,c)$ is a unitary\footnote{
  See Sec.\ \ref{sec:Review} for the meaning of unitarity.}
string field carrying $\Ngh=0$.
For $(N+1)$-brane solution, we took a special $U$ specified by
$N+1$ real parameters $\alpha_k$ ($k=0,1,\cdots,N$), among which only
$[N/2]$ are independent.\footnote{
$[x]$ denotes the greatest integer less than or equal to $x$.
}
Our $U$ is in fact a natural extension of that appearing in the
former construction of the tachyon vacuum ($N=-1$) and the 2-brane
($N=1$) solutions \cite{Okawa,ES09,MS1,HKwn,MS2,HKinfty}. 

Though our candidate solution \eqref{eq:Psi=UQBU^-1} is of pure-gauge,
it is a non-trivial matter whether and in what sense it satisfies the
EOM \eqref{eq:EOM}. This is because of the singularity of $\Psi$
\eqref{eq:Psi=UQBU^-1} coming from the zero-eigenvalue of the string
field $K$ as a $*$-multiplication operator.
For properly treating the singularity at $K=0$, we adopted 
the $\Ke$-regularization \cite{HKwn} of replacing $K$ by
$\Ke\equiv K+\veps$ with $\veps$ being a positive infinitesimal
constant.
For any string field $\calO(K,B,c)$, we define $\calO_\veps$ by
\begin{equation}
\calO_\veps\equiv\calO\bigr|_{K\to \Ke}=\calO(\Ke,B,c) .
\end{equation}
Then, as conditions on the solution and hence on the parameters
$\alpha_k$, we took the following two.
One is that the EOM test of the $\Ke$-regularized solution
$\Psie$,
\begin{equation}
\Psie=(U\QB U^{-1})_\veps ,
\label{eq:Psie}
\end{equation}
against the solution itself,
\begin{equation}
\calT=\int\!\Psie*\left(\QB\Psie+\Psie^2\right) ,
\label{eq:calT}
\end{equation}
should vanish, namely, $\calT=0$.
Due to the $\Ke$-regularization, the EOM of $\Psie$ is
violated apparently by $O(\veps)$,
$\QB\Psie+\Psie^2=O(\veps)$,
which, together with the singularity of $O(1/\veps)$ at $K=0$, can
make $\calT$ non-vanishing in the limit $\veps\to +0$.
Another condition on the $(N+1)$-brane solutions is that the
``winding number'' $\calN$ defined by
\begin{equation}
\calN=\frac{\pi^2}{3}\int\!\Psie^3 ,
\label{eq:calN}
\end{equation}
should be equal to the integer $N$.
If $\calT=0$ holds,
the number of D25-branes determined from the action is equal to
$\calN+1$.
The reason why we call $\calN$ \eqref{eq:calN} ``winding number''
is due to its analogy to the winding number
\begin{equation}
\calW[g]=\frac{1}{24\pi^2}\int_M\tr\left(gd g^{-1}\right)^3 ,
\label{eq:calW}
\end{equation}
of the mapping $g(x)$ from a three-manifold $M$ to a Lie group.
This analogy was emphasized and examined in \cite{HKwn}.
Understanding the topological structure of CSFT through this analogy
was one of the motivations of \cite{ACMBSol}.

Summarizing, our conditions on the $(N+1)$-brane solution are the
following two:
\begin{equation}
\calN=N,\qquad
\calT=0 .
\label{eq:2conds_Sol}
\end{equation}
We examined the two conditions \eqref{eq:2conds_Sol} for various
integers $N=2,3,\cdots, 35$, and found that there indeed exists the
parameter set $\setalpha$ satisfying the two conditions for a
respective $N$.
More precisely, we found, as far as we have tested for various $N$, that
$\calN$ and $\calT$ are given in the following form in terms of
linear functions $f_m(\alpha_k)$ ($m=1,2,\cdots,[N/2]$):
\begin{equation}
\calN=N+\sum_{m=1}^{[N/2]}f_m(\alpha_k)\,z^{2m} ,
\qquad
\calT=-\sum_{m=1}^{[N/2]}t_m f_m(\alpha_k)\,z^{2(m-1)} ,
\label{eq:calN_calT_by_fn}
\end{equation}
where $t_m$ are numerical coefficients and $z$ is
$z\equiv 2\pi i$.
For $N=2,3,4,5$, for which the number of independent $\alpha_k$,
namely, $[N/2]$, is less than or equal to two, the two conditions
\eqref{eq:cond_alpha} uniquely determines $\setalpha$.
On the other hand, for $N\ge 6$, we found that $\setalpha$ is uniquely
determined by imposing stronger conditions
\begin{equation}
f_m(\alpha_k)=0
\qquad
\left(m=1,2,\cdots,[N/2]\right) .
\label{eq:fm=0}
\end{equation}
Note that the number of conditions \eqref{eq:fm=0} is equal to that of
the independent components of $\setalpha$.

As we described in detail in \cite{ACMBSol},  our solutions are not
complete ones since they fail to satisfy the EOM test against the Fock
states. This problem might be resolved by some kind of improvements of
the solutions, or by some consistent truncation of the space of
fluctuations around multi-branes which excludes the Fock states.
However, even if this problem remains, our construction which realizes
integer values for $\calN$ is expected to give some hint on
understanding the meaning of $\calN$ as ``winding umber'', and
furthermore, the topological aspects in CSFT.

The purpose of this paper is to determine the parameter set
$\setalpha$ of our solution for a generic integer $N$ and thereby
complete our construction of multi-brane solutions.
(In \cite{ACMBSol}, we confirmed only that $\setalpha$
satisfying the conditions \eqref{eq:fm=0} exists
for sample values of $N$ ($=2,3,\cdots,35$).)
Namely, we first show that eq.\ \eqref{eq:calN_calT_by_fn} for $\calN$
and $\calT$ is valid for a generic $N$ and give linear functions
$f_m(\alpha_k)$ in closed forms. 
Then, we solve \eqref{eq:fm=0} for $\setalpha$ to find that each
$\alpha_k$ is given in a closed form by using the Bernoulli numbers
(see \eqref{eq:alpha_k_by_B_n}).\footnote{
The Bernoulli numbers also appear in the tachyon vacuum solution of
Schnabl \cite{Schnabl}, though there is no direct relevance to the
present case.
}

Besides giving $\setalpha$ for a generic $N$, we also report some
findings concerning our construction.
First, we present our analysis of the brane number of
our solutions determined from their gravitational coupling.
In \cite{ACMBSol}, we made a guess that the stronger conditions
\eqref{eq:fm=0} on $\setalpha$ would be obtained by considering
various physical demands in addition to the two in
\eqref{eq:2conds_Sol}, and the brane number from the gravitational
coupling would be one of such demands.
In this paper, we find, contrary to our expectation, that the brane
number calculated from the gravitational coupling for our candidate
$(N+1)$-brane solution is completely independent of $\setalpha$ and is
equal to $N+1$ without any anomalous contributions.

Secondly, we show that the unitary string field $U$
in \eqref{eq:Psi=UQBU^-1} has a simple expression as an exponential
function; the exponent is given by $\CRBc\equiv\CR{B}{c}$
multiplied by functions of $K$.
Though at present we do not have any concrete application of this
fact, we hope that it would give a clue to the understanding of
$\calN$ \eqref{eq:calN} as a ``winding number''.

The rest of this paper is organized as follows.
In Sec.\ 2, we review the construction of \cite{ACMBSol}
outlined above in more detail by using the convenient notation adopted
in \cite{ACMBSol}.
Then, in Sec.\ 3, we present the linear functions $f_m(\alpha_k)$ and
the solution $\setalpha$ to $f_m(\alpha_k)=0$ both in closed forms
for a generic $N$.
In Secs.\ 4 and 5, we discuss the gravitational coupling of our
multi-brane solutions, and the unitary string field $U$ as an
exponential function, respectively.
The final section (Sec.\ 6) is devoted to a summary and discussions.
In the Appendix, we present a proof of eq.\
\eqref{eq:calN_calT_by_fn}.

\section{More details of the construction}
\label{sec:Review}

In the Introduction, we summarized the construction of multi-brane
solutions proposed in \cite{ACMBSol}.
In this section, we review the construction in more detail for use
in later sections.

In our construction, we use the concise notation for products of
string fields.
In this notation, we attach to any string field $\calO$ a pair of
integers $(a,b)$ specifying the left and the right of $\calO$ to write
$\calO_{ab}$. As for the elements $(K,B,c)$ of the $\KBc$
algebra which are also string fields, we attach to $c$ a pair $(a,b)$
to write $c_{ab}$, while to $K$ and $B$, which are commutative with
each other, we assign a single number to write $K_a$ and $B_a$.
For example, let us consider the string field $\calO$ with $\Ngh=1$ of
the form
\begin{equation}
\calO=f_1(K)\,c\,f_2(K)\,Bc\,f_3(K)+g_1(K)\,c\,g_2(K)\,Bc\,g_3(K) .
\end{equation}
This is expressed in our notation as
\begin{equation}
\calO_{13}=A_{123}\,c_{12}(Bc)_{23} ,
\end{equation}
where $A_{123}$ consisting only of $K$ is given by
\begin{equation}
A_{123}=f_1(K_1)f_2(K_2)f_3(K_3)+g_1(K_1)g_2(K_2)g_3(K_3) ,
\end{equation}
and $(Bc)_{ab}=B_a c_{ab}$.
In the present notation, we are allowed to put functions of $K$ such
as $A_{123}$ at any place without any ambiguity.

Our construction of multi-brane solutions starts with the most
generic form of $U$ which consists only of $(K,B,c)$, and is unitary,
$U^\dg=U^{-1}$, and real.\footnote{
As in \cite{ACMBSol}, $\dg$ denotes the composition of the BPZ and the
hermitian conjugation.
The reality of $U$ means that $U(K,B,c)$ does not contain
any imaginary unit $i$.
}
The unitarity of $U$ is for the self-conjugateness $\Psi^\dg=\Psi$ of
$\Psi$ given by \eqref{eq:Psi=UQBU^-1}, and the reality of $U$ is
simply for the sake of simplicity.
Let us express $U$ with $\Ngh=0$ generically as
\begin{equation}
U_{ab}=\frac{1}{\Gm_a}\,\id_{ab}-\frac{\pF_{ab}}{\Gm_b}(Bc)_{ab} ,
\label{eq:U_ab=}
\end{equation}
where $\id$ is the identity string field, and
$\Gm_a$ and $\pF_{ab}$ are given in terms of real functions
$\Gm(x)$ and $\pF(x,y)$ as
\begin{equation}
\Gm_a=\Gm(K_a),
\qquad
\pF_{ab}=\pF(K_a,K_b) .
\label{eq:Gma=Gm(Ka)_pFab=pF(Ka,Kb)}
\end{equation}
Then, the unitarity of $U$ is realized if $(\Gm_a,\pF_{ab})$ satisfies
\begin{equation}
\pF_{ab}=\pF_{ba},
\qquad
\pF_{aa}=1-(\Gm_a)^2,
\label{eq:cond_pFab_Gma}
\end{equation}
namely, $\pF(x,y)=\pF(y,x)$ and $\pF(x,x)=1-\Gm(x)^2$.
If the conditions \eqref{eq:cond_pFab_Gma} hold,
$U^{-1}$ is given by
\begin{equation}
(U^{-1})_{ab}=(U^\dg)_{ab}
=\Gm_a\id_{ab}+\frac{\pF_{ab}}{\Gm_a}(Bc)_{ab} ,
\end{equation}
and the pure-gauge configuration \eqref{eq:Psi=UQBU^-1} reads
\begin{equation}
\Psi_{ac}=(U\QB U^{-1})_{ac}=E_{abc}(cK)_{ab}(Bc)_{bc} ,
\label{eq:Psi_ab}
\end{equation}
with $(cK)_{ab}=c_{ab}K_b$ and
\begin{equation}
E_{abc}=\pF_{ac}+\pF_{ab}\,\frac{1}{\Gm_b^2}\,\pF_{bc} .
\label{eq:E_abc}
\end{equation}
Upon the $\Ke$-regularization,
the EOM of $\Psie$ is violated by $O(\veps)$,
\begin{equation}
\left(\QB\Psie+\Psie^2\right)_{ac}
=\veps\times \left(E_{abc}\right)_\veps(c\Ke)_{ab}\,c_{bc} .
\end{equation}

Construction of solutions carrying a generic integer $\calN$ and
satisfying the EOM test $\calT=0$ is not an easy
matter.
Solutions with $\calN=\pm 1$ representing the 2-brane
and the tachyon vacuum
\cite{Okawa,ES09,MS1,HKwn,MS2,HKinfty}
correspond to the following $(\Gm_a,\pF_{ab})$:
\begin{equation}
\Gm_a=\sqrt{G_a} ,
\qquad
\pF_{ab}=\sqrt{\left(1-G_a\right)\left(1-G_b\right)} ,
\label{eq:(Gm_a,pF_ab)_N=1}
\end{equation}
given in terms of $G_a\equiv G(K_a)$ with a simple pole/zero at $K=0$;
$G(K)=O\bigl(K^{\mp 1}\bigr)$.
This solution also satisfies the EOM test $\calT=0$.
However, the extension of this type of solution by adopting $G(K)$
with higher order pole/zero at $K=0$ leads to non-integer $\calN$ and
the violation of the EOM test \cite{MS1,HKwn,MS2,HKinfty}.

Then, we proposed in \cite{ACMBSol} that a $(N+1)$-brane solution
($N=1,2,3,\cdots$) satisfying the two conditions of
\eqref{eq:2conds_Sol} can be realized by adopting the following
$(\Gm_a,\pF_{ab})$:
\begin{align}
\Gm_a&=G_a^{N/2} ,
\\
\pF_{ab}&=\prod_{k=0}^N\left(1-G_a^k G_b^{N-k}\right)^{\alpha_k}
=-\prod_{k=0}^N\left(G_a^k G_b^{N-k}-1\right)^{\alpha_k} ,
\end{align}
where $\alpha_k$ ($k=0,1,\cdots,N$) are real parameters satisfying
\begin{equation}
\alpha_{N-k}=\alpha_k ,
\qquad
\sum_{k=0}^N\alpha_k=1 ,
\label{eq:cond_alpha}
\end{equation}
and $G(K)$ here should have a simple pole at $K=0$ and no other
zeros/poles in the complex half-plane $\Re K\ge 0$
including $K=\infty$, but is otherwise arbitrary.
A typical example is
\begin{equation}
G(K)=\frac{1+K}{K} .
\label{eq:G(K)=(1+K)/K}
\end{equation}
Note that the two relations in \eqref{eq:cond_alpha} correspond to
those in \eqref{eq:cond_pFab_Gma}, and the number of
independent $\alpha_k$ is $[N/2]$.
The parameter set $\setalpha$ should be chosen in such a way that
the two conditions \eqref{eq:2conds_Sol} are satisfied.

\begin{table}[htbp]
\centering
\renewcommand{\arraystretch}{1.2}
\begin{tabular}{|c|c|}
\hline
$N$& $\alpha_0, \alpha_1,\cdots, \alpha_{[N/2]-1},\alpha_{[N/2]}$
\\
\hline\hline
$1$ & $1/2$
\\ \hline
$2$ & $0,\, 1$
\\ \hline
$3$ & $-1/6,\, 2/3$
\\ \hline
$4$ & $0,\, -1/2,\, 2$
\\ \hline
$5$ & $1/6,\, -1,\, 4/3$
\\ \hline
$6$ & $0,\, 2/3,\, -4,\, 23/3$
\\ \hline
$7$ & $-3/10,\, 12/5,\, -32/5,\, 24/5$
\\ \hline
$8$ & $0,\, -3/2,\, 12,\, -37,\, 54$
\\ \hline
$9$ & $5/6,\, -25/3,\, 32,\, -56,\, 32$
\\ \hline
$10$ & $0,\, 5,\, -50,\, 417/2,\, -468,\, 610$
\\ \hline
$11$ & $-691/210,\, 1382/35,\, -20528/105,\, 10652/21,\,
-24384/35,\, 12224/35$
\\ \hline
\end{tabular}
\caption{$\setalpha$ determined from \eqref{eq:fm=0} for
  $N=1,2,\cdots,11$.
  Here, we show only the independent components $\alpha_k$
  ($k=0,1,\cdots,[N/2]-1$) and, in addition, $\alpha_{[N/2]}$.
}
\label{tab:N_alpha_k}
\end{table}

In \cite{ACMBSol}, we gave $\calN[\alpha_k]$ and $\calT[\alpha_k]$ as
functions of $\alpha_k$. However, since they were given in complicated
forms, we could not solve \eqref{eq:2conds_Sol} for $\setalpha$ for a
generic $N$. Instead, we examined the two conditions of
\eqref{eq:2conds_Sol} for various sample values of $N$
($=2,3,\cdots,35$) to find that $\setalpha$ satisfying the conditions
indeed exists.
Table \ref{tab:N_alpha_k} shows $\setalpha$ which fulfills the two
conditions of \eqref{eq:2conds_Sol} for $N=1,2,\cdots, 11$, as
examples. A number of explanations are necessary.
First, for $N=1$, we uniquely have $(\alpha_0,\alpha_1)=(1/2,1/2)$
from \eqref{eq:cond_alpha}, reproducing \eqref{eq:(Gm_a,pF_ab)_N=1}.
Secondly, we found, as far as we have tested for various $N$, that
$\calN$ and $\calT$ are given in the form of
\eqref{eq:calN_calT_by_fn} in terms of
linear functions $f_m(\alpha_k)$ ($m=1,2,\cdots,[N/2]$).
The values of $\setalpha$ for $N\ge 2$ in the table are particular
solutions to \eqref{eq:2conds_Sol} determined by imposing the
conditions \eqref{eq:fm=0} (which are stronger than the original
\eqref{eq:2conds_Sol} in the cases $N\ge 6$).

\section{$\setalpha$ for a generic $N$}
\label{sec:alpha}

In this section, we present the main results of this paper.
First, we can show, using the complicated expressions of
$\calN$ and $\calT$ given in \cite{ACMBSol}, that
eq.\ \eqref{eq:calN_calT_by_fn} for them is in fact true for a generic
$N$ and that $f_m(\alpha_k)$ and $t_m$ there are given by
\begin{align}
f_m(\alpha_k)&=\frac{1}{(2m-2)!}\sum_{k=2m}^N
\Pmatrix{k+1\\ 2m+1}\alpha_k
=\frac{1}{(2m-2)!}\sum_{k=0}^{N-2m}
\Pmatrix{N-k+1\\ 2m+1}\alpha_k ,
\label{eq:f_m(alpha_k)}
\\
t_m&=\frac{4\left(2m+1\right)}{2m-1} .
\label{eq:t_m}
\end{align}
We outline the derivation of these results, which is straightforward
but lengthy, in Appendix \ref{app:calN_calT}.
Of course, eqs.\ \eqref{eq:f_m(alpha_k)} and \eqref{eq:t_m} reproduce
$f_m(\alpha_k)$ and $t_m$ given in Sec.\ 5 of \cite{ACMBSol} for
$N=2,\cdots, 7$ and $11$.

Next, let us consider obtaining $\setalpha$ which satisfies the
stronger conditions \eqref{eq:fm=0} for a generic $N$.
For this purpose, we introduce the function $F(t)$ defined by
\begin{equation}
F(t)=\sum_{k=0}^N\alpha_k\,t^{k+1} ,
\label{eq:F(t)}
\end{equation}
by which $f_m(\alpha_k)$ \eqref{eq:f_m(alpha_k)} is expressed as
\begin{equation}
f_m(\alpha_k)=\frac{1}{(2m-2)!\,(2m+1)!}
\left.\drv{^{2m+1} F(t)}{t^{2m+1}}\right|_{t=1} .
\label{eq:fm_by_F(t)}
\end{equation}
Then, the problem of finding $\setalpha$ satisfying \eqref{eq:fm=0}
as well as \eqref{eq:cond_alpha} is equivalent to finding
$F(t)$ which is a polynomial in $t$ of order $N+1$, and satisfies the
following conditions:
\begin{align}
F(t)&=t^{N+2} F(1/t) ,
\label{eq:F(t)=t^N+2F(1/t)}
\\
F(1)&=1 ,
\label{eq:F(1)=1}
\\[3mm]
\left.\drv{^{2m+1} F(t)}{t^{2m+1}}\right|_{t=1}&=0
\qquad\left(m=1,2,\cdots,[N/2]\right) ,
\label{eq:d^2m+1F(t)/dt^2m+1=0}
\end{align}
which correspond to the first and the second conditions in
\eqref{eq:cond_alpha}, and \eqref{eq:fm=0}, respectively.
The polynomial $F(t)$ lacks the constant term and satisfies
$F(0)=0$. However, we do not need to take care of this fact as another
condition on $F(t)$ since it is an automatic consequence of
\eqref{eq:F(t)=t^N+2F(1/t)} and the fact that
$F(1/t)=O\!\left(1/t^{N+1}\right)$ as $t\to 0$.

For constructing such $F(t)$, it is convenient to reexpress $F(t)$ as a
polynomial in $1-t$:
\begin{equation}
F(t)=C_0+\sum_{n=1}^{N+1}C_n\left(1-t\right)^n .
\label{eq:F(t)_by_C_n}
\end{equation}
Then, the two conditions \eqref{eq:F(1)=1} and
\eqref{eq:d^2m+1F(t)/dt^2m+1=0} are restated as
\begin{equation}
C_0=1 ,
\label{eq:C_0=1}
\end{equation}
and
\begin{equation}
C_{2m+1}=0
\qquad\left(m=1,2,\cdots,[N/2]\right) ,
\label{eq:C_odd>1=0}
\end{equation}
respectively.
As for another condition \eqref{eq:F(t)=t^N+2F(1/t)}, from
\begin{align}
&t^{N+2}F(1/t)
=\sum_{k=0}^{N+1}(-1)^k C_k\,t^{N+2-k}\left(1-t\right)^k
\nn\\
&=\sum_{n=0}^{N+1}(-1)^n\left[
\sum_{k=0}^nC_k\Pmatrix{N+2-k\\ n-k}\right]
\left(1-t\right)^n+(-1)^{N+2}\left(\sum_{k=0}^{N+1}C_k\right)
\left(1-t\right)^{N+2} ,
\end{align}
we obtain the following two kinds of conditions for $C_n$:
\begin{align}
\sum_{k=0}^n\Pmatrix{N+2-k\\ n-k}C_k&=(-1)^n C_n
\qquad
\left(n=0,1,2,\cdots,N+1\right) ,
\label{eq:condCone}
\\
\sum_{k=0}^{N+1}C_k&=0 .
\label{eq:condCtwo}
\end{align}
Note that \eqref{eq:condCtwo} is equivalent to $F(0)=0$.

Let us see how eqs.\ \eqref{eq:condCone} and \eqref{eq:condCtwo}
combined with \eqref{eq:C_0=1} and \eqref{eq:C_odd>1=0} determine
$C_n$.
Eq.\ \eqref{eq:condCone} with $n=0$ is trivially satisfied.
This equation with $n=1$ together with \eqref{eq:C_0=1} gives
$C_1=-N/2-1$.
Then, using \eqref{eq:C_odd>1=0}, namely, $C_{n=\textrm{odd}\ge 3}=0$,
eqs.\ \eqref{eq:condCone} with $n=2,\cdots,N+1$ and \eqref{eq:condCtwo}
are expressed unifiedly as
\begin{equation}
\sum_{k=0}^{n-1}\Pmatrix{N+2-k\\ n-k}C_k=0
\qquad
\left(n=2,3,\cdots,N+2\right) ,
\label{eq:codCone+two}
\end{equation}
or, equivalently,
\begin{equation}
C_n=-\frac{1}{N+2-n}\sum_{k=0}^{n-1}
\Pmatrix{N+2-k\\ n+1-k}C_k
\qquad\left(n=1,2,\cdots,N+1\right) .
\label{eq:Recur_C_n}
\end{equation}
Starting with $C_0=1$ \eqref{eq:C_0=1}, the recursion equation
\eqref{eq:Recur_C_n} successively determines $C_n$.
For example, the first ten $C_n$ are given as follows:
\begin{align}
C_1&=-\frac12\Pmatrix{N+2\\ 1},
\quad
C_2=\frac16\Pmatrix{N+2\\ 2},
\quad
C_3=0,
\quad
C_4=-\frac{1}{30}\Pmatrix{N+2\\ 4},
\quad
C_5=0,
\nn\\
C_6&=\frac{1}{42}\Pmatrix{N+2\\ 6},
\quad
C_7=0,
\quad
C_8=-\frac{1}{30}\Pmatrix{N+2\\ 8},
\quad
C_9=0,
\quad
C_{10}=\frac{5}{66}\Pmatrix{N+2\\ 10} ,
\label{eq:sample_C_n}
\end{align}
where $C_1$ agrees with what we have already obtained, and
$C_{n=\textrm{odd}\ge 3}=0$ are not assumed but they are
consequences of \eqref{eq:Recur_C_n}.
The ten $C_n$ given in \eqref{eq:sample_C_n} take a special form.
First, the $N$ dependence of $C_n$ is through the binomial coefficient
$\sPmatrix{N+2\\ n}$. Second, the rational number multiplying the
binomial coefficient in $C_n$ is the $n$-th Bernoulli number. 
To show that these two are general properties valid for all $n$, let
us express $C_n$ as
\begin{equation}
C_n=\Pmatrix{N+2\\ n}B_n .
\label{eq:C_n_by_B_n}
\end{equation}
Then, the initial condition \eqref{eq:C_0=1} and the recursion
equation \eqref{eq:Recur_C_n} for $C_n$ are rewritten
into $N$-independent equations for $B_n$:
\begin{equation}
B_0=1,
\qquad
B_n=-\frac{1}{n+1}\sum_{k=0}^{n-1}\Pmatrix{n+1\\ k}B_k
\quad
\left(n\ge 1\right) ,
\label{eq:B_0=1_Recur_B_n}
\end{equation}
or, equivalently,
\begin{equation}
\sum_{k=0}^n\Pmatrix{n+1\\ k}B_k=\delta_{n,0}
\qquad (n\ge 0) .
\label{eq:RecRelB_n}
\end{equation}
Eq.\ \eqref{eq:B_0=1_Recur_B_n} (or \eqref{eq:RecRelB_n}) are nothing
but one of the defining equations of the Bernoulli numbers $B_n$.
Examples of non-vanishing $B_n$ are
\begin{equation}
B_1=-\frac12,\ B_2=\frac16,\ B_4=-\frac{1}{30},\ B_6=\frac{1}{42},\
B_8=-\frac{1}{30},\ B_{10}=\frac{5}{66},\ B_{12}=-\frac{691}{2730} .
\label{eq:exampleB_n}
\end{equation}
We also have
\begin{equation}
B_{2m+1}=0\qquad (m\ge 1) .
\label{eq:B_2m+1=0}
\end{equation}

Having determined the coefficients $C_n$ in \eqref{eq:F(t)_by_C_n},
$\alpha_k$ is obtained by reexpanding \eqref{eq:F(t)_by_C_n} around
$t=0$:
\begin{equation}
\alpha_k=(-1)^{k+1}\sum_{n=k+1}^{N+1}\Pmatrix{n\\ k+1}C_n
=(-1)^{k+1}\Pmatrix{N+2\\ k+1}
\sum_{n=k+1}^{N+1}\Pmatrix{N-k+1\\ n-k-1}B_{n} .
\label{eq:alpha_k_by_B_n}
\end{equation}
This is our desired expression of $\alpha_k$ for a generic $N$.
Of course, it reproduces $\alpha_k$ given in Table \ref{tab:N_alpha_k}
for $N=1,2,\cdots,11$.

The Bernoulli numbers also appear in the tachyon vacuum solution of
Schnabl \cite{Schnabl}, though there is no direct relation between
the two cases. It would be interesting if there exists some
mathematical reason for the appearance of the Bernoulli numbers in
classical solutions of CSFT.

\section{Energy from the gravitational  coupling}

In this section, we consider the energy density of our candidate
solution specified by $\setalpha$ from its coupling with graviton
\cite{AHandNI,GRSZ,Ellwood,KKT}.
The correspondent of $\calN$ \eqref{eq:calN} defined by the
gravitational coupling is
\begin{equation}
\calNgrav=2\pi^2\int\!\Vmid\,\Psie ,
\label{eq:calNgrav}
\end{equation}
where $\Vmid$ is the zero-momentum graviton vertex at the string
midpoint, and is given in the sliver frame by
\begin{equation}
\Vmid=\frac{2}{\pi i}\,c\p X(i\infty)c\p X(-i\infty) .
\label{eq:Vmid}
\end{equation}
Our interest is what kind of condition on $\setalpha$ is obtained by
demanding $\calNgrav=N$ for our candidate $(N+1)$-brane solution.

Recall that the present solution is given by
\eqref{eq:Psi_ab} with $E_{abc}$ of \eqref{eq:E_abc}.
Here, we define the function $E(x,y,w)$ of the three variables
$(x,y,w)$ by
\begin{equation}
E_{abc}=E(K_a,K_b,K_c) ,
\end{equation}
and introduce the inverse Laplace transform $\wt{E}(p,q,r)$
of $y E(x,y,w)$:
\begin{equation}
y E(x,y,w)=\int_0^\infty\!dp\,e^{-px}\int_0^\infty\!dq\,e^{-qy}
\int_0^\infty\!dr\,e^{-rw}\,\wt{E}(p,q,r) .
\end{equation}
Then, the $\Ke$-regularized solution $\Psie$ is expressed (in the
conventional notation) as
\begin{equation}
\Psie=\int_0^\infty\!dp\int_0^\infty\!dq\int_0^\infty\!dr
\,\wt{E}(p,q,r)\,e^{-p\Ke}\,c\,e^{-q\Ke} Bc\,e^{-r\Ke} .
\label{eq:Psie_by_wtE}
\end{equation}
Using \eqref{eq:Vmid} and \eqref{eq:Psie_by_wtE} and introducing the
regularization $\Lambda$ for the string midpoint, $\calNgrav$
\eqref{eq:calNgrav} is calculated as follow:
\begin{align}
\calNgrav
&=2\pi^2\int_0^\infty\!dp\int_0^\infty\!dq\int_0^\infty\!dr
\,e^{-\veps(p+q+r)}\,\wt{E}(p,q,r)
\nn\\
&\qquad\times\frac{2}{\pi i}
\VEV{Bc(0) c(r+i\Ld)(\p X)(r+i\Ld)
  c(r-i\Ld)(\p X)(r-i\Ld) c(r+p)}_{\ell=p+q+r}
\nn\\
&=-\int_0^\infty\!dp\int_0^\infty\!dq\int_0^\infty\!dr
\,e^{-\veps(p+q+r)}
\left(r+p\right)\wt{E}(p,q,r)
\nn\\
&=\left.\left(\Drv{}{x}+\Drv{}{w}\right)
y\,E(x,y,w)\right|_{x=y=w=\veps} ,
\label{eq:calc1_calNgrav}
\end{align}
where $\VEV{\cdots}_\ell$ is the correlator on the cylinder of
circumference $\ell$.
In \eqref{eq:calc1_calNgrav}, we have taken the limit
$\Lambda\to\infty$ with $\veps$ kept non-zero by using the following
expressions of the matter and the ghost parts of the correlator in the
approximation of $\Lambda/\ell \gg 1$:\footnote{
  See Appendix A of \cite{HataKojitaGrav} for the correlators on the
  sliver frame in the present convention.}
\begin{align}
&\VEV{\p X(r+i\Ld) \p X(r-i\Ld)}_{\ell}
=\frac{\pi^2}{2\ell^2}
\left(\sin\frac{2\pi i\Ld}{\ell}\right)^{-2}
\simeq
-\frac{2\pi^2}{\ell^2}\,e^{-4\pi\Ld/\ell} ,
\\[3mm]
&\VEV{Bc(0)c(r+i\Ld)c(r-i\Ld)c(r+p)}_{\ell}
=\frac{\ell^2}{\pi^3}\biggl\{
(r+p)\sin\frac{\pi(r+i\Ld)}{\ell}\sin\frac{\pi(r-i\Ld)}{\ell}
\sin\frac{2\pi i\Ld}{\ell}
\nn\\
&\qquad
+\sum_\pm(\pm)(r\mp i\Ld)
\sin\frac{\pi(r\pm i\Ld)}{\ell}\sin\frac{\pi(r+p)}{\ell}
\sin\frac{\pi(p\mp i\Ld)}{\ell}
\biggr\}
\simeq\frac{i\ell^2}{8\pi^3}\left(r+p\right)e^{4\pi\Ld/\ell} .
\end{align}
Then, the calculation \eqref{eq:calc1_calNgrav} of $\calNgrav$ is
completed by using that $E(x,y,w)$ is given by
\begin{equation}
E(x,y,w)
=\pF(x,w)+\frac{\pF(x,y)\pF(y,w)}{G(y)^N} ,
\end{equation}
with
\begin{equation}
\pF(x,y)=\prod_{k=0}^N\left(1-G(x)^k G(y)^{N-k}\right)^{\alpha_k} ,
\end{equation}
and that
\begin{equation}
\left.\Drv{}{x}\pF(x,y)\right|_{y=x}
=-\sum_{k=0}^N k\alpha_k\,G'(x) G(x)^{N-1} .
\end{equation}
We obtain
\begin{equation}
\calNgrav=2\veps\left.\Drv{}{x}\pF(x,y)\right|_{x=y=\veps}
\left(1+\frac{\pF(\veps,\veps)}{G(\veps)^N}\right)
=-2\sum_{k=0}^Nk\alpha_k\times\veps\frac{G'(\veps)}{G(\veps)}
=N ,
\label{eq:calNgrav=N}
\end{equation}
where we have used the identity
\begin{equation}
\sum_{k=0}^N k\alpha_k=\frac{N}{2} ,
\label{eq:sum_k_kalpha_k=N/2}
\end{equation}
which is a consequence of \eqref{eq:cond_alpha}, and the fact that
$G(z)$ has a simple pole at $z=0$.
Namely, the number of branes of our candidate solution determined by
the gravitational coupling is equal to the desired value $N+1$ for any
$\setalpha$.

In \cite{ACMBSol} and in Sec.\ \ref{sec:alpha} of this paper,
we determined $\alpha_k$ by imposing $[N/2]$ conditions
\eqref{eq:fm=0} which are stronger than the original two of
\eqref{eq:2conds_Sol} for $N\ge 6$. Our expectation in \cite{ACMBSol}
was that the conditions of \eqref{eq:fm=0} would be obtained by
considering other physical requirements than \eqref{eq:2conds_Sol},
and that $\calNgrav=N$ would be one of such requirements.
Our finding in this section is that, contrary to our expectation,
the energy density determined by the gravitational coupling gives no
constraint on $\setalpha$.

\section{$U$ as an exponential function}
\label{sec:U=exp}

In this section, we show that our unitary string field $U$
\eqref{eq:U_ab=} given by $(\Gm_a,\pF_{ab})$ satisfying
\eqref{eq:cond_pFab_Gma} has an expression as a simple exponential
function.
We have in mind applying this expression of $U$ to the study of
the meaning of $\calN$ \eqref{eq:calN} as a ``winding number''.

We show that $U$ has the following expression:
\begin{equation}
U=\exp\calA=\id+\sum_{n=1}^\infty\frac{1}{n!}\calA^n ,
\label{eq:U=expcalA}
\end{equation}
where $\calA$ is given by
\begin{equation}
\calA_{ab}=\calH_{ab}\CRBc_{ab} ,
\end{equation}
in terms of $\CRBc$ and $\calH(K)$ defined by
\begin{equation}
\CRBc=\CR{B}{c}=2Bc-\id ,
\label{eq:CRBc}
\end{equation}
and
\begin{equation}
\calH_{ab}=\frac12\,\frac{\pF_{ab}}{1-\Gm_a\Gm_ b}
\,\ln\left(\Gm_a\Gm_b\right) .
\label{eq:calH}
\end{equation}
Note that $\CRBc$ is anti-self-conjugate and its square is the
identity:
\begin{equation}
\CRBc^\dg=-\CRBc,
\qquad
\CRBc^2=\id .
\label{eq:prop_CRBc}
\end{equation}
The properties of $\calH$ are
\begin{equation}
\calH_{ab}=\calH_{ba},
\qquad
\calH_{aa}=\ln\Gm_a .
\label{eq:prop_calH}
\end{equation}
Then, the anti-self-conjugateness of $\CRBc$ together with the
symmetric and real property of $\calH_{ab}$ implies that $\calA$ is
anti-self-conjugate:
\begin{equation}
\left(\calA^\dg\right)_{ab}
=\calH_{ba}\left(\CRBc^\dg\right)_{ab}
=\calH_{ab}\left(-\CRBc\right)_{ab}=-\calA_{ab}
\quad\Rightarrow\quad
\calA^\dg=-\calA .
\end{equation}

Expression \eqref{eq:U=expcalA} of $U$ can be derived by taking
the logarithm of $U$ given by \eqref{eq:U_ab=}.
Here, instead, we show that \eqref{eq:U=expcalA} reproduces
\eqref{eq:U_ab=}.
For this purpose, we express $\calA^n$ ($n=1,2,3,\cdots$) as a sum of
terms linear in $\CRBc$ and $\id$.
First, for $n=2$ we obtain
\begin{align}
(\calA^2)_{ab}&=\calA_{a1}\calA_{1b}
=\calH_{a1}\calH_{1b}\,\CRBc_{a1}\CRBc_{1b}
=\calH_{a1}\calH_{1b}\left(
\id_{a1}\CRBc_{1b}-\CRBc_{a1}\id_{1b}+\id_{a1}\id_{1b}\right)
\nn\\
&=\left(\calH_{aa}-\calH_{bb}\right)\calA_{ab}
+(\calH_{aa})^2\,\id_{ab} ,
\label{eq:calA^2}
\end{align}
where we have used the formula
\begin{equation}
\CRBc_{ab}\CRBc_{bc}=\id_{ab}\CRBc_{bc}-\CRBc_{ab}\id_{bc}
+\id_{ab}\id_{bc} .
\label{eq:CRBcCRBc}
\end{equation}
Note that \eqref{eq:CRBcCRBc} is reduced to $\CRBc^2=\id$ in
\eqref{eq:prop_CRBc} in the special case where there is no $K_b$
multiplying \eqref{eq:CRBcCRBc}.
Then, $\calA^3$ and $\calA^4$ are successively calculated to give
\begin{align}
(\calA^3)_{ab}&=
\left(\calH_{aa}^2-\calH_{aa}\calH_{bb}+\calH_{bb}^2\right)\calA_{ab} ,
\\
(\calA^4)_{ab}&=
\left(\calH_{aa}^3-\calH_{aa}^2\calH_{bb}+\calH_{aa}\calH_{bb}^2
-\calH_{bb}^3\right)\calA_{ab}+\calH_{aa}^4\id_{ab} .
\end{align}
From the above examples, we guess that $\calA^n$ for a generic integer
$n\;(\ge 1)$ is of the following form:
\begin{align}
(\calA^n)_{ab}
&=\sum_{k=0}^{n-1}\calH_{aa}^{n-1-k}\left(-\calH_{bb}\right)^k
\calA_{ab}
+\begin{cases}
\calH_{aa}^n\id_{ab} & (n:\;\textrm{even})
\\[2mm]
0 & (n:\;\textrm{odd})
\end{cases}
\nn\\
&=\frac{\calH_{aa}^n-(-\calH_{bb})^n}{\calH_{aa}+\calH_{bb}}
\,\calA_{ab}
+\frac{1+(-1)^n}{2}\,\calH_{aa}^n\id_{ab} .
\label{eq:calA^n_gen_n}
\end{align}
In fact, the validity of \eqref{eq:calA^n_gen_n} is proved by
induction by using \eqref{eq:CRBcCRBc}.
Note that the last expression of \eqref{eq:calA^n_gen_n} formally
applies to the case of $n=0$.

Now, using \eqref{eq:calA^n_gen_n}, we find that
\eqref{eq:U=expcalA} for $U$ agrees with its original expression
\eqref{eq:U_ab=}:
\begin{align}
U_{ab}&=\left(\exp\calA\right)_{ab}
=\sum_{n=0}^\infty\frac{1}{n!}(\calA^n)_{ab}
=\frac{e^{\calH_{aa}}-e^{-\calH_{bb}}}{\calH_{aa}+\calH_{bb}}
\,\calA_{ab}
+\frac{e^{\calH_{aa}}+e^{-\calH_{bb}}}{2}\,\id_{ab}
\nn\\
&=\frac12\left(\Gm_a+\frac{1}{\Gm_a}\right)\id_{ab}
-\frac{\pF_{ab}}{2\Gm_b}\,\CRBc_{ab} ,
\end{align}
where we have used \eqref{eq:calH} and \eqref{eq:prop_calH}.

The expression \eqref{eq:U=expcalA} of $U$ in terms of $\CRBc$
with properties \eqref{eq:prop_CRBc} suggests its correspondence
to
\begin{equation}
g(\bm{x})=\exp\left(f(r)\,T\right)
\qquad\Bigl(T\equiv\frac{i\bm{x}\cdot\bm{\tau}}{r},
\quad r\equiv\abs{\bm{x}}\Bigr) ,
\label{eq:g(x)}
\end{equation}
which is a mapping from $\bm{x}\in\mathbb{R}^3$ to $SU(2)\simeq S^3$.
Since $T$ satisfies
\begin{equation}
T^\dagger =-T,\qquad T^2=-\id_2 ,
\label{eq:T}
\end{equation}
it is guessed that there might exist some kind of correspondence
between $\left(\CRBc,\calH\right)$ for $U$ and $\left(T,f(r)\right)$
for $g(\bm{x})$. Furthermore, the winding number $\calW$
\eqref{eq:calW} of the mapping $g(\bm{x})$ is given by
$\calW=\left(f(\infty)-f(0)\right)/\pi$, which becomes an integer if
$g(\bm{x})$ is regular at $r=0$ and $\infty$, namely, if $f(0)$ and
$f(\infty)$ are integer multiples of $\pi$.
It would be interesting if similar regularity arguments on $U$ in
terms of $\calH$ lead to the conditions \eqref{eq:fm=0} on
$\setalpha$.

\section{Summary and discussions}

In this paper, we have determined the parameters $\alpha_k$ specifying
the analytic $(N+1)$-brane solution proposed in \cite{ACMBSol}.
For this, we first showed, for a generic $N$, that
the ``winding number'' $\calN$ \eqref{eq:calN} and the EOM test
$\calT$ \eqref{eq:calT} are given in the form of
\eqref{eq:calN_calT_by_fn}, and identified the linear functions
$f_m(\alpha_k)$ appearing there.
Then, we solved the conditions $f_m(\alpha_k)=0$ to get a particular
solution $\setalpha$ to $\calN=N$ and $\calT=0$ in a closed form.
We found that each $\alpha_k$ is given by using the Bernoulli
numbers.
The obtained results for a generic $N$ reproduce those given in
\cite{ACMBSol} for special values of $N$.
We also presented our results on the energy density of the solution
defined by its gravitational coupling and on the expression of the
unitary string field $U$ constituting the solution as an exponential
function. In particular, we found that the energy density from the
gravitational coupling is totally independent of $\setalpha$ and is
equal to the correct value.

Our remaining tasks are as follows.
First, we would like to understand the form of our solution specified
by the obtained $\setalpha$ \eqref{eq:alpha_k_by_B_n} from the view
point of the ``regularity'' of the solution or of the unitary string
field $U$ (possibly at $K=0$), and furthermore, understand the meaning
of $\calN$ \eqref{eq:calN} as a ``winding number''.
For this, the expression of $U$ as an exponential function given in
Sec.\ \ref{sec:U=exp} and its similarity to the unitary matrix
$g(\bm{x})$ \eqref{eq:g(x)} could give some hints.
Of course, it is interesting if we could find a profound reason
why the Bernoulli numbers appear in $\alpha_k$.

Secondly, we would like to understand what is the most general
$\setalpha$ allowed for our solution.
In this paper, we obtained a particular solution $\setalpha$ to
the two conditions $\calN=N$ and $\calT=0$ in \eqref{eq:2conds_Sol} by
solving (generically) stronger conditions $f_m(\alpha_k)=0$
\eqref{eq:fm=0}.
Our problem is whether only $\setalpha$ satisfying \eqref{eq:fm=0}
is allowed, or a general solutions to \eqref{eq:2conds_Sol}
with $[N/2]-2$ arbitrary parameters (in the case $N\ge 6$) are
all allowed and the arbitrary parameters represent the freedom of
continuous gauge equivalent deformation of the solution.
For this, we have to clarify the meaning of the ``allowed solution''.
This could be ``regular'' solutions in the sense mentioned above,
or solutions satisfying physical requirements other than the two of 
\eqref{eq:2conds_Sol}.
In relation to this, the problem of the failure of the EOM test of the
present solution against the Fock states \cite{ACMBSol} needs serious
study. This should include the analysis of fluctuation modes around
our solution (cf.\ \cite{HataFluct}).

\section*{Acknowledgments}

We would like to thank T.\ Kojita, H.\ Ohki, S.\ Seki, T.\ Takahashi
for valuable discussions.

\appendix

\section{Proof of \eqref{eq:calN_calT_by_fn}}
\label{app:calN_calT}

In this appendix, we show that $\calN$ and $\calT$ are given by
\eqref{eq:calN_calT_by_fn} in terms of linear functions
$f_m(\alpha_k)$ \eqref{eq:f_m(alpha_k)} and the coefficients $t_m$ 
\eqref{eq:t_m}.

\subsection{Derivation of $\calN$ in \eqref{eq:calN_calT_by_fn}}

In \cite{ACMBSol}, we presented the following expression of $\calN$
\eqref{eq:calN} obtained by using the fact that the variation of
$\calN$ with respect to $U$ is formally the CSFT-integration of a
BRST-exact quantity:
\begin{align}
\calN&=-\sum_{k,\ell=0}^N\alpha_k\alpha_\ell\biggl[
\ell\Bigl(\calN_{k-\ell+1,-k,\ell}
+\calN_{\ell,-k,k-\ell+1}-\calN_{\ell-k,0,k-\ell+1}
-\calN_{\ell,0,-\ell+1}\Bigr)
\nn\\
&\qquad
+\left(N-\ell\right)\Bigl(
\calN_{k-\ell,-k,\ell+1}
+\calN_{\ell+1,-k,k-\ell}
-\calN_{\ell-k+1,0,k-\ell}
-\calN_{\ell+1,0,-\ell}\Bigr)\biggr] ,
\label{eq:calN_by_calNn1n2n3}
\end{align}
where $\calN_{n_1,n_2,n_3}$ ($n_1+n_2+n_3=1$) is given by
\begin{equation}
\calN_{n_1,n_2,n_3}=S_{n_1,n_2,n_3}
-S_{n_1,n_2-1,n_3}-S_{n_1,n_2,n_3-1}+S_{n_1,n_2-1,n_3-1} ,
\label{eq:calN_n1n2n3=sumS}
\end{equation}
in terms of $S_{m_1,m_2,m_3}$ which is defined by
\begin{equation}
\frac{1}{\pi^2}S_{m_1,m_2,m_3}
=\veps\int_{u_0}^0\!du\,u^{1+\sum_{a=1}^3m_a}\,
\int\!c\,G_u^{m_1}c\,G_u^{m_2}c\,G_u^{m_3} .
\label{eq:S_m1m2m3=}
\end{equation}
In \eqref{eq:S_m1m2m3=}, $u_0$ is an arbitrary positive constant, and
$G_u$ with parameter $u$ is
\begin{equation}
G_u=\frac{1}{\Ke+u} .
\label{eq:G_u}
\end{equation}
Explicit expression of $S_{m_1,m_2,m_3}$ is given in \cite{ACMBSol}
(eqs.\ (3.26)--(3.29)).
Here, we instead use another expression:\footnote{
The following relation holds between $f_{P,Q}$ appearing in
$S_{m_1,m_2,m_3}$ (3.28) of \cite{ACMBSol} and the present
$\calI_{m,n}$:
$$
m\,f_{m+1,n}-n\,f_{n+1,m}=-\frac{(m+n+1)!}{4\pi}\,\calI_{m,n} .
$$
More directly, \eqref{eq:S_m1m2m3_by_calI} is derived by using
the following $(s,z)$-integration expression \cite{MS1,MS2}:
$$
\int\!c\,G_u^{m_1}c\,G_u^{m_2}c\,G_u^{m_3}
=\int_0^\infty\!ds\,\frac{s^2}{(2\pi)^3\,i}
\int_{-i\infty}^{i\infty}\!\frac{dz}{2\pi i}\,e^{sz}\left[
G_u^{m_1}\Ds G_u^{m_2+m_3}+G_u^{m_2}\Ds G_u^{m_3+m_1}
+G_u^{m_3}\Ds G_u^{m_1+m_2}\right] .
$$
}
\begin{equation}
S_{m_1,m_2,m_3}=-\frac{\left(1+\sum_{a=1}^3m_a\right)!}{4\pi}
\left(\calI_{m_1,m_2+m_3}+\calI_{m_2,m_3+m_1}+\calI_{m_3,m_1+m_2}\right) ,
\label{eq:S_m1m2m3_by_calI}
\end{equation}
where $\calI_{m,n}$ is defined by
\begin{align}
\calI_{m,n}&=\Im\Res_{z=0,2\pi i}\frac{e^z}{z^m\left(z-2\pi i\right)^n}
\nn\\
&=\theta(n\ge 1)\sum_{k=0}^{n-1}
\Pmatrix{-m\\ n-1-k}\frac{\Im\left(2\pi i\right)^{k-m-n+1}}{k!}
-\left(m\rightleftarrows n\right) .
\label{eq:calI_mn=}
\end{align}
Note that $\calI_{m,n}$ is anti-symmetric:
\begin{equation}
\calI_{m,n}=-\calI_{n,m} .
\label{eq:antisym_calI}
\end{equation}

The sum of the indices $m_a$ of $S_{m_1,m_2,m_3}$
appearing in \eqref{eq:calN_n1n2n3=sumS} is either of the following
three:
\begin{equation}
M\equiv\sum_{a=1}^3 m_a=1,0, -1 .
\label{eq:M=sum_ma}
\end{equation}
We calculate the contribution of the $M=1$, $0$ and $-1$ terms,
namely, the first term, the second and the third terms, and the last
term, of $\calN_{n_1,n_2,n_3}$ \eqref{eq:calN_n1n2n3=sumS} to $\calN$
separately. In this calculation, we repeatedly use
\eqref{eq:cond_alpha}, the anti-symmetry \eqref{eq:antisym_calI}, and
that $S_{m_1,m_2,m_3}$ is totally symmetric with respect to its
indices and vanishes if at least one of the three $m_a$ is equal to
zero.

\noindent
\underline{$M=1$ term}

The contribution of the $M=1$ term in $\calN_{n_1,n_2,n_3}$
\eqref{eq:calN_n1n2n3=sumS},
\begin{equation}
\calN^{(1)}_{n_1,n_2,n_3}
=S_{n_1,n_2,n_3}
=-\frac{1}{2\pi}\left(\calI_{n_1,1-n_1}
+\calI_{n_2,1-n_2}+\calI_{n_3,1-n_3}\right) ,
\end{equation}
to $\calN$ \eqref{eq:calN_by_calNn1n2n3} is given by
\begin{align}
\calN^{(1)}&=-2\sum_{k,\ell=0}^N\alpha_k\alpha_\ell\left[
\ell\,S_{k-\ell+1,-k,\ell}+\left(N-\ell\right)S_{k-\ell,-k,\ell+1}\right]
\nn\\
&=\frac{1}{\pi}\sum_{k,\ell=0}^N\alpha_k\alpha_\ell\Bigl[
\ell\left(\calI_{k-\ell+1,\ell-k}+\calI_{-k,1+k}+\calI_{\ell,1-\ell}\right)
+\left(N-\ell\right)\left(
\calI_{k-\ell,1-k+\ell}+\calI_{-k,1+k}+\calI_{\ell+1,-\ell}\right)
\Bigr]
\nn\\
&=\frac{1}{\pi}\sum_{k,\ell=0}^N\alpha_k\alpha_\ell\Bigl[
\ell \left(\calI_{k-\ell+1,\ell-k}+\calI_{\ell-k,k-\ell+1}\right)
+N\left(\calI_{-k,1+k}+\calI_{\ell+1,-\ell}\right)
+\ell\left(\calI_{\ell,1-\ell}-\calI_{\ell+1,-\ell}\right)
\Bigr]
\nn\\
&=\frac{1}{\pi}\sum_{\ell=1}^N\ell\alpha_\ell
\left(\calI_{\ell,1-\ell}-\calI_{\ell+1,-\ell}\right) ,
\label{eq:calN^(1)}
\end{align}
where we have made the replacement of the summation indices
$(k,\ell)\to(N-k,N-\ell)$ for the $\calI_{k-\ell,1-k+\ell}$ term on
the second line to obtain the $\calI_{\ell-k,k-\ell+1}$ term 
on the third line,
and then used the anti-symmetry \eqref{eq:antisym_calI}
to reach the final expression.

\noindent
\underline{$M=0$ term}

The contribution of the $M=0$ term,
\begin{equation}
\calN^{(0)}_{n_1,n_2,n_3}
=-S_{n_1,n_2-1,n_3}-S_{n_1,n_2,n_3-1} ,
\end{equation}
to $\calN$ is calculated similarly to the $M=1$ case:
\begin{align}
\calN^{(0)}&=\sum_{k,\ell=0}^N\alpha_k\alpha_\ell\Bigl[
\ell\bigl(2S_{k-\ell+1,-k-1,\ell}+S_{k-\ell+1,-k,\ell-1}
+S_{\ell,-k,k-\ell}
-S_{\ell-k,-1,k-\ell+1}-S_{\ell,-1,-\ell+1}\bigr)
\nn\\
&\quad
+\left(N-\ell\right)\bigl(
2S_{k-\ell,-k-1,\ell+1}+S_{k-\ell,-k,\ell}
+S_{\ell+1,-k,k-\ell-1}
-S_{\ell-k+1,-1,k-\ell}-S_{\ell+1,-1,-\ell}\bigr)\Bigr]
\nn\\
&=-\frac{1}{2\pi}\sum_{\ell=0}^N\ell\alpha_\ell
\left(\calI_{\ell-1,1-\ell}-\calI_{\ell+1,-\ell-1}\right)
+\frac{N}{2\pi}\,\calI_{-1,1} .
\label{eq:calN^(0)}
\end{align}
In obtaining this, we have used, in particular, that
$\sum_{k,\ell}\alpha_k\alpha_\ell\calI_{k-\ell,\ell-k}=0$,
which is due to the anti-symmetry \eqref{eq:antisym_calI}.
The last term $\left(N/(2\pi)\right)\calI_{-1,1}$ is from the four
$S_{m_1,m_2,m_3}$ with $m_2=-1$.

\noindent
\underline{$M=-1$ term}

Finally, the contribution of the $M=-1$ term,
\begin{equation}
\calN^{(-1)}_{n_1,n_2,n_3}=S_{n_1,n_2-1,n_3-1} ,
\end{equation}
is given by
\begin{align}
\calN^{(-1)}&=-\sum_{k,\ell=0}^N\alpha_k\alpha_\ell\Bigl[
\ell\left(S_{k-\ell+1,-k-1,\ell-1}
+S_{\ell,-k-1,k-\ell}-S_{\ell-k,-1,k-\ell}-S_{\ell,-1,-\ell}\right)
\nn\\
&\qquad
+\left(N-\ell\right)\left(
S_{k-\ell,-k-1,\ell}+S_{\ell+1,-k-1,k-\ell-1}
-S_{\ell-k+1,-1,k-\ell-1}-S_{\ell+1,-1,-\ell-1}\right)\Bigr]
\nn\\
&=\frac{1}{2\pi}\sum_{\ell=0}^N\ell\alpha_\ell
\left(\calI_{\ell-1,-\ell}-\calI_{\ell,-\ell-1}\right)
-\frac{N}{2\pi}\,\calI_{-1,0} .
\label{eq:calN^(-1)}
\end{align}

\noindent
\underline{Total of $\calN$}

Summing the above three results, \eqref{eq:calN^(1)},
\eqref{eq:calN^(0)} and \eqref{eq:calN^(-1)},
we find that the whole of $\calN$ is
given by
\begin{align}
\calN&=\calN^{(1)}+\calN^{(0)}+\calN^{(-1)}
\nn\\
&=\frac{N}{2\pi}\left(\calI_{-1,1}-\calI_{-1,0}\right)
\nn\\
&\quad
+\frac{1}{2\pi}\sum_{\ell=0}^N \ell\alpha_\ell\left[
2\left(\calI_{\ell,1-\ell}-\calI_{\ell+1,-\ell}\right)
-\left(\calI_{\ell-1,1-\ell}-\calI_{\ell+1,-\ell-1}\right)
+\left(\calI_{\ell-1,-\ell}-\calI_{\ell,-\ell-1}\right)\right]
\nn\\
&=N+\frac{1}{2\pi}\Im\sum_{\ell=2}^N\alpha_\ell
\sum_{k=3}^{\ell+1}\frac{1}{(k-3)!}\Pmatrix{\ell+1\\ k}
\left(2\pi i\right)^k .
\label{eq:calN_total}
\end{align}
In deriving the last expression, we have used \eqref{eq:calI_mn=} for
$\calI_{m,n}$, and, in particular, that
\begin{equation}
\calI_{-1,1}=2\pi,\qquad \calI_{-1,0}=0 .
\end{equation}
Since the $k$-summation in \eqref{eq:calN_total} is restricted to odd
integers, we can express $k$ as $k=2m+1$ and exchange the order of the
$\ell$- and $m$-summations. This leads to $\calN$ in
\eqref{eq:calN_calT_by_fn} with $f_m(\alpha_k)$ given by
\eqref{eq:f_m(alpha_k)}.

\subsection{Derivation of $\calT$ in \eqref{eq:calN_calT_by_fn}}

In \cite{ACMBSol}, we showed that the EOM test $\calT$
\eqref{eq:calT} is given by
\begin{equation}
\calT=\sum_{k,\ell=0}^N\alpha_k\alpha_\ell\Bigl[
\calT_{k-\ell,0,\ell-k,0}-2\,\calT_{k,0,\ell-k,-\ell}
-2\,\calT_{k-\ell,-k,\ell,0}+2\,\calT_{k,-k,\ell,-\ell}
+2\,\calT_{k-\ell,-k,N,\ell-N}
\Bigr] ,
\label{eq:calT_by_calTn1n2n3n4}
\end{equation}
where $\calT_{n_1,n_2,n_3,n_4}$ with $\sum_{a=1}^4n_a=0$ is defined by
\begin{equation}
\calT_{n_1,n_2,n_3,n_4}
=\veps\int\!Bc\frac{1}{\Ke^{n_3}}c\frac{1}{\Ke^{n_4-1}}
c\frac{1}{\Ke^{n_1}}c\frac{1}{\Ke^{n_2-1}} .
\label{eq:calTn1n2n3n4}
\end{equation}
In \eqref{eq:calT_by_calTn1n2n3n4}, we have used the
$n_1\rightleftarrows n_3$ exchange symmetry,
\begin{equation}
\calT_{n_1,n_2,n_3,n_4}=\calT_{n_3,n_2,n_1,n_4} ,
\end{equation}
to simplify the original expression (4.9) in \cite{ACMBSol}.
Further, $\calT_{n_1,n_2,n_3,n_4}$ is given explicitly by
\begin{align}
\calT_{n_1,n_2,n_3,n_4}&=n_1\left(h_{n_3+n_4-1}-h_{n_3}-h_{n_4-1}\right)
+n_3\left(h_{n_4+n_1-1}-h_{n_4-1}- h_{n_1}\right)
\nn\\
&\quad
+\left(n_4-1\right)
\left(h_{n_3+n_4}+h_{n_4+n_1}-h_{n_4}-h_{-n_2}\right) ,
\end{align}
in terms of $h_n$ defined by
\begin{equation}
h_n=\frac{1}{\pi}\Im\left[
\theta(n\le -2)\sum_{k=0}^{-n-2}\Pmatrix{-n\\ k+2}
\frac{\left(2\pi i\right)^k}{k!}
-\theta(n\ge 1)\sum_{k=0}^{n-1}\Pmatrix{n+1\\ k+2}
\frac{\left(2\pi i\right)^k}{k!}\right] .
\end{equation}
Note that, by definition, $h_n$ satisfies
\begin{equation}
h_{-n-1}=-h_n .
\label{eq:h_n-1=-h_n}
\end{equation}

Let us obtain a simpler expression of each of the five terms on the
RHS of \eqref{eq:calT_by_calTn1n2n3n4} in terms of $h_n$.
The first term is calculated as follows:
\begin{align}
\sum_{k,\ell=0}^N\alpha_k\alpha_\ell\calT_{k-\ell,0,\ell-k,0}
&=\sum_{k,\ell=0}^N\alpha_k\alpha_\ell\bigl[
(k-\ell)\left(h_{\ell-k-1}-h_{\ell-k}-h_{-1}\right)
\nn\\
&\qquad
+(\ell-k)\left(h_{k-\ell-1}-h_{-1}-h_{k-\ell}\right)
-\left(h_{\ell-k}+h_{k-\ell}-2h_0\right)\bigr]
\nn\\
&=-2\sum_{k,\ell=0}^N\alpha_k\alpha_\ell h_{k-\ell} ,
\end{align}
where we have used
\begin{equation}
h_0=h_{-1}=0 ,
\label{eq:h_0=h_-1=0}
\end{equation}
and the relation $h_{\ell-k-1}=-h_{k-\ell}$ due to
\eqref{eq:h_n-1=-h_n} to obtain the last expression.
Other four terms in \eqref{eq:calT_by_calTn1n2n3n4} are simplified in
a similar manner:
\begin{align}
-2\sum_{k,\ell}\alpha_k\alpha_\ell\,\calT_{k,0,\ell-k,-\ell}
&=-2\sum_{k,\ell}\alpha_k\alpha_\ell\,\calT_{k-\ell,-k,\ell,0}
\nn\\
&=\sum_k\left(N-2k\right)\alpha_k\left(h_k+h_{-k}\right)
+2\sum_{k,\ell}\left(k+\ell+1\right)\alpha_k\alpha_\ell h_{k-\ell} ,
\nn\\
2\sum_{k,\ell}\alpha_k\alpha_\ell\,\calT_{k,-k,\ell,-\ell}
&=2\sum_k\left(k+1\right)\alpha_k\left(h_k+h_{-k}\right)
-2\sum_{k,\ell}\left(k+\ell+1\right)\alpha_k\alpha_\ell h_{k-\ell} ,
\nn\\
2\sum_{k,\ell}\alpha_k\alpha_\ell\,\calT_{k-\ell,-k,N,\ell-N}
&=2\sum_k\left(2k-N\right)\alpha_k\left(h_k+h_{-k}\right)
-2N\sum_{k,\ell}\alpha_k\alpha_\ell h_{k-\ell} .
\end{align}
In this derivation, we have used \eqref{eq:sum_k_kalpha_k=N/2},
and have made the changes the of summation indices
$k\rightleftarrows\ell$ and $(k,\ell)\to(N-k,N-\ell)$ if necessary.
Then, summing the five contributions, we find that the whole of
$\calT$ is given by
\begin{align}
\calT&=2\sum_{k,\ell=0}^N\alpha_k\alpha_\ell
\left(k+\ell-N\right)h_{k-\ell}
+2\sum_{\ell=0}^N\left(\ell+1\right)\alpha_\ell
\left(h_\ell+h_{-\ell}\right)
\nn\\
&=-\frac{2}{\pi}\Im\sum_{\ell=2}^N\alpha_\ell\left(\ell+1\right)
\sum_{k=1}^{\ell-1}\Pmatrix{\ell\\ k+1}
\frac{(2\pi i)^k}{k!} .
\label{eq:calT_total}
\end{align}
Note that the first term after the first equality vanishes as seen by
making the change of the summation indices
$(k,\ell)\to(N-\ell,N-k)$.
Then, expressing $k$ in \eqref{eq:calT_total} as $k=2m-1$ and
exchanging the order of the $\ell$- and $m$-summations, we obtain
$\calT$ in \eqref{eq:calN_calT_by_fn} with $f_m(\alpha_k)$ and $t_m$
given by \eqref{eq:f_m(alpha_k)} and \eqref{eq:t_m}, respectively.


\end{document}